\documentclass[showpacs,amsmath,aps,superscriptaddress]{revtex4}
\usepackage{txfonts}
\usepackage{graphicx,amssymb,mathrsfs,bm,color,times}
\usepackage{float}

\newcommand{\be}{\begin{equation}}
\newcommand{\ee}{\end{equation}}
\newcommand{\bea}{\begin{eqnarray}}
\newcommand{\eea}{\end{eqnarray}}
\newcommand{\bsube}{\begin{subequations}}
\newcommand{\esube}{\end{subequations}}

\newcommand{\Eq}[1]{Eq.\,(\ref{#1})}

\newcommand{\etal}{{\it et al.\ }}

\newcommand{\la}{\langle}
\newcommand{\ra}{\rangle}

\newcommand{\beq}{\begin{equation}}
\newcommand{\eeq}{\end{equation}}
\newcommand{\beqn}{\begin{eqnarray}}
\newcommand{\eeqn}{\end{eqnarray}}
\newcommand{\bsub}{\begin{subequations}}
\newcommand{\esub}{\end{subequations}}

\newcommand{\Rmnum}[1]{\uppercase\expandafter{\romannumeral #1}}

\begin{document}
\title { Low-field electron mobility of InSb nanowires:
Numerical efforts to larger cross sections}

\author{Wei Feng}
\email{fwphy@tju.edu.cn}
\affiliation{Department of Physics, Tianjin University,
Tianjin 300072, China}

\author{Chen Peng}
\affiliation{Department of Physics, Renmin University of China,
Beijing 100872, China}

\author{Shuang Li}
\affiliation{Center for Advanced Quantum Studies and
Department of Physics, Beijing Normal University,
Beijing 100875, China}

\author{Xin-Qi Li}
\email{lixinqi@bnu.edu.cn}
\affiliation{Center for Advanced Quantum Studies and
Department of Physics, Beijing Normal University,
Beijing 100875, China}

\date{\today}

\begin{abstract}
Within the framework of Boltzmann equation, we present a
$\mathbf{k\cdot p}$ theory based study for the low-field
mobilities of InSb nanowires (InSb NWs) with relatively
large cross sectional sizes (with diameters up to 51.8 nm).
For such type of large size nanowires,
the intersubband electron-phonon scattering
is of crucial importance to affect the scattering rate and then
the mobility. In our simulation, the lowest 15 electron subbands
and 50 transverse modes of phonons are carefully accounted for.
We find that, up to the 51.84 nm diameter,
the mobility monotonously increases with the diameter,
not yet showing any saturated behavior.
We also find that, while the bulk InSb mobility
is considerably higher than the bulk Si,
the small size (e.g. $\sim 3$ nm diameter) nanowires
from both materials have similar magnitude of mobilities.
This implies, importantly, that
the mobility of the InSb NWs would decrease faster than
the SiNWs as we reduce the cross sectional size of the nanowires.
\end{abstract}

\maketitle

{\flushleft Mobility is one}
of the most important figures
to affect material properties and device applications.
For nano-electronic purpose, device miniaturization
usually requires engineering the bulk materials
to smaller sizes or to lower dimensions.
In this context, how the mobility changes with the size
and dimensionality is a crucial problem.

For theoretical investigations, the first intrinsic mechanism
of mobility is the electron phonon scattering.
This requires to know the full knowledge of electron and phonon states,
and to properly account for the summation
over a large number of discrete phonon modes and electron subbands
under the restriction of energy conservation.
We notice that so far only very limited
theoretical calculations have been reported.
Among them,
representative examples are the mobility calculations
of Si nanowires (SiNWs), up to diameters of a few nanometers
\cite{Ana08j,Ana08n,Fon06,Mic93,San93,San92}.

In the past years the InSb material and InSb NWs \cite{Gul15,For08}
have attracted intensive attention and promise sound applications
in high-speed field effect and new-concept quantum devices,
owing to the high mobility (up to 77000 ${\rm cm}^2 {\rm V}^{-1}
{\rm s}^{-1}$ at room temperature) and large Land\'e $g$ factor.
A distinguished example may be the demonstration
of Majorana fermion in the hybrid system by contacting
the InSb NW with a superconductor \cite{Kou12,Xu12,Rok12,Das12}.

In this work we present a state-of-the-art calculation
for the phonon limited electron mobility of the InSb NWs.
In particular, we make efforts to extend theoretical studies
to nanowires with relatively large diameters,
for instance, up to 51.8 nm.
For this purpose, instead of the first-principle
or the tight-banding schemes,
we employ the eight-band $\mathbf{k\cdot p}$
theory to calculate the electronic structures of the InSb NWs,
with tolerable computational expenses 
\cite{Ost96,Kol03,Mal96,Sip94,Fore97,Bah90,Pok01,Yang06,Nov05}.
The $\mathbf{k\cdot p}$ theory should be appropriate
for the {\it low-field mobility} calculation,
for which the acoustic phonon scattering
is also the dominant {\it intrinsic mechanism}.
For the phonon modes, we employ the continuum media model.
We apply the deformation-potential theory to address
the electron-phonon interaction and apply the Boltzmann
equation to formulate the low-field mobility calculation.  \\
\\
\\
{\bf\large  Results} \\
\\
{\bf Electron Energy Subbands.}
We employ the eight-band $\mathbf{k\cdot p}$ theory
for the calculation of electronic states,
which can provide reliable result around
a given high-symmetry point $\mathbf{k_0}$
(usually the $\Gamma$ point).
In this method, one conventionally denotes the eight {\it known} states
at $\mathbf{k_0}$ by $|J,J_z\ra\equiv |n\ra$,
where $J$ and $J_z$ are the total angular momentum
and its components which characterize, respectively,
the conduction-band electron ($J=1/2$ and $J_z=\pm 1/2$),
the valence-band heavy ($J_z=\pm 3/2$)
and light ($J_z=\pm 1/2$) holes with $J=3/2$,
and the spin-orbit coupling split states in valence band
with $J_z=\pm 1/2$.

To apply the $\mathbf{k\cdot p}$ theory to confined system as
the nanowire with circular cross section, it would be convenient
to use the eigen-functions solved from a cylindrical wire
to constitute the representation basis of the envelope function,
which plays a role to modulate
the periodic kernel function of the bulk Bloch state.
This basis function,
$\psi_{L,m}(k_z,\mathbf{r})=\la \mathbf{r}|k_z,L,m\ra$,
is simply given by
\begin{equation}\label{eq-wf}
\psi_{L,m}(k_z,\mathbf{r})=\frac{1}{\sqrt{l}}e^{ik_zz}\frac{1}
{\sqrt{\pi}}\frac{1}{RJ_{L+1}(\alpha^{L}_m)}
J_L(\frac{\alpha^L_m}{R}r)   \,e^{iL\varphi} \,,
\end{equation}
where $l$ and $R$ are respectively the length
and cross section radius of the nanowire.
$k_z$ is the wave vector along the $z$-axis.
The lateral quantization of the wave function
is characterized the quantum numbers $L$ and $m$.
The $J_L(x)$ is the $L$-th order Bessel function
of the first kind, with $\alpha^L_m$ of its $m$-th root.
Here we have denoted the polar coordinates
in the circular cross section by $(r,\varphi)$.

Using the basis set $\{ |n\ra\otimes |k_z,L,m\ra  \}$,
the electron eigen-energies and states can be solved from
\bea \label{Hnn}
\sum_{n',L',m'}
\la k_z,L,m| {\cal H}_{n,n'} |k_z,L',m'\ra
C^{(\nu)}_{n',L',m'}
= E_{\nu}(k_z) C^{(\nu)}_{n,L,m} \,,
\eea
where ${\cal H}_{n,n'}$ is the $\mathbf{k\cdot p}$
Hamiltonian matrix element of bulk material.
Owing to lateral confinement,
the components of the wave vector
$k_x$ and $k_y$ in ${\cal H}_{n,n'}$
become now the momentum operators,
acting on the spatial coordinates
of the lateral wavefunction ($\psi_{L,m}$)
in the cross section of the nanowire.
Corresponding to the eigen-energy $E_{\nu}(k_z)$,
the eigen-state reads
$|\Psi_{\nu}\ra = \sum_{n,L,m} \, C^{(\nu)}_{n,L,m}
\,|n\ra \otimes |k_z,L,m\ra $,
which will be used in this work to calculate the
electron-phonon scattering rate.

In the $\mathbf{k\cdot p}$ theory,
there exits the so-called {\it spurious solution} problem,
caused by the {\it incompleteness} of the basis functions
included in practical computation.
A couple of schemes were proposed to partially
overcome this difficulty, such as discarding terms
in the Hamiltonian \cite{Kol03} or rejecting the
unphysical large $k$ solutions \cite{Sip94,Mal96}.
In this work we adopt the method proposed in Ref.\cite{Fore97},
by modifying parameters to make the $k^2$ terms be zero
in the conduction-band matrix elements
in ${\cal H}_{n,n'}$, meanwhile properly
fitting the conduction-band effective mass.
After this type of treatment/modification to
the Hamiltonian matrix, no spurious solution will appear
and one is able to recover all the other effective masses
obtained by experiment.

Applying the eight-band $\mathbf{k\cdot p}$ theory
to the zinc-blende InSb \cite{Bah90}
and using the material parameters from Ref.\ \cite{Vur01},
we are able to calculate the electronic states
of the InSb nanowires up to relatively large sizes .
In Fig.\ 1(a) we show the result of a couple
of conduction subbands in the [001] direction,
for an InSb NW with diameter 51.8 nm.
We see that with the increase of the size of the nanowire,
the energy spacing between the subbands decreases.
This will make the intersubband scattering be of
crucial importance in the mobility calculation
for large size nanowires. In this work we will
make particular efforts to account for this issue. \\
\\
{\bf Phonon Spectrum.}
For the acoustic phonons, we apply the continuous medium model
to solve the vibrational modes
under a freestanding boundary condition (FSBC).
After some algebras, the problem is reduced to
the following Pochhammer-Chree equation \cite{Dan00}
\begin{align}\label{P-C}
&\frac{2q_l}{R}(q_t^2+q_z^2)J_1(q_l R)J_1(q_t R)
-(q_t^2-q_z^2)^2J_0(q_l R)J_1(q_t R)  \\ \nonumber
&-4q_z^2q_l q_tJ_1(q_l R)J_0(q_t R)=0  \,.
\end{align}
$J_{0}$ and $J_{1}$ are the Bessel functions.
$q_l$ and $q_t$ are defined through
$q_{l,t}^2=\frac{\omega^2}{v_{l,t}^2}-q_z^2$,
where $\omega$ and $q_z$ are respectively the vibration frequency
and the longitudinal wave vector
(along the axial direction of the nanowire),
while $v_l$ and $v_t$ are the speeds
of the longitudinal and transverse waves of bulk material.
For InSb in the [100] direction, we take
$v_l=3.4\times 10^5\; {\rm cm/s}$ and
$v_t=2.3\times 10^5\; {\rm cm/s}$,
following Ref.\ \cite{web}.

In Fig.\ 1(b), we display the dispersion relation
of the lowest 50 transverse phonon modes,
for an InSb NW with diameter of 51.8 nm.
This number of transverse modes will be included
into the electron phonon scattering
in our mobility calculation.
In particular, we remark here that the {\it circular}
cross sectional nanowires under FSBC can support
the existence of transverse vibrational mode with zero frequency.
This property will have dramatic effect on the
momentum relaxation rate of electrons. \\
\\

\begin{figure}[!htp]
\includegraphics[scale=0.65]  {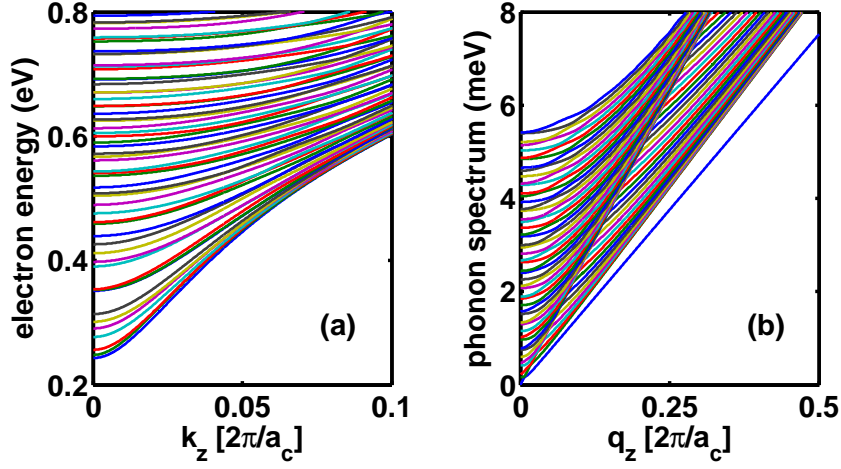}  
\caption{
(a)
Conduction subbands of InSb nanowire along the
[001] direction with diameter 51.8 nm,
based on the $\mathbf{k\cdot p}$ theory calculation.
(b)
Phonon spectrum of the same InSb nanowire
calculated from a continuum media model.
In both plots, $a_c$ is the lattice constant.  }
\end{figure}

{\flushleft\bf Momentum Relaxation Rate.}
For the dominant intrinsic mechanism at low fields,
we employ the deformation potential (DP) model
for the electron-acoustic-phonon scattering \cite{Zhang10,Mur10}.
The Hamiltonian reads
$H_{ep}=E_a\nabla \cdot \mathbf{u} $,
where $\mathbf{u}$ is the lattice displacement (deformation)
and $E_a$ is the DP constant.
For InSb, $E_a=5.08$ eV, from Ref.\  \cite{Vur01}.
Applying the Fermi's golden rule, one can derive
an expression for the momentum relaxation rate
(see the `Method' part for some details).
In numerical implementation,
particular attention should be paid
to the large number of mode summations
under the restriction of energy conservation.
We show in Fig.\ 2 several representative results.

For thin nanowires with small cross section,
it is a good approximation to consider electrons
populated mainly on the lowest subband
and the electron phonon scattering dominantly
within the same subband (i.e., intrasubband scattering).
In Fig.\ 2(a) we exemplify the result
of an InSb nanowire with radius 1.3 nm.
In this plot
we separately present the rates from phonon emission
and absorption, and find that only phonon emission peaks appear
in the rate with the increase of the initial energy of the electron.
Each peak indicates that a more transverse mode
is newly involved into the scattering.
Notably, the `half peak' near zero frequency is resulted from
scattering with the transverse zero-mode mentioned above in Fig.\ 1(b).

In Fig.\ 2(b) we plot the summed total relaxation rate
caused by phonon emission and absorption,
for a couple of cross sectional sizes.
As already revealed from Fig.\ 2(a), we know that
the peaks in the total rate are resulted from phonon emissions.
With the increase of the nanowire sizes,
the peaks move to lower frequencies,
while at the same time the scattering rate is reduced,
owing to the lower energies of the phonon modes, being thus
{\it less efficient} to take away the energy of the electron.

From simple consideration, we expect that the intersubband scattering
will become important with the size increase of the nanowires.
Indeed, as shown in Fig.\ 2(c) for an InSb NW with radius 5.2 nm,
the effect of inclusion of the intersubband scattering
(solid red curve) is prominent for higher electron energy
(i.e. over 30 meV), as manifested here in both the magnitude
change and additional new peaks.
This result clearly tells us that, for larger size nanowires,
one must taken into account the inter-subband scattering
in the mobility calculation. \\
\\

\begin{figure}[!htp]
\includegraphics[scale=0.7] {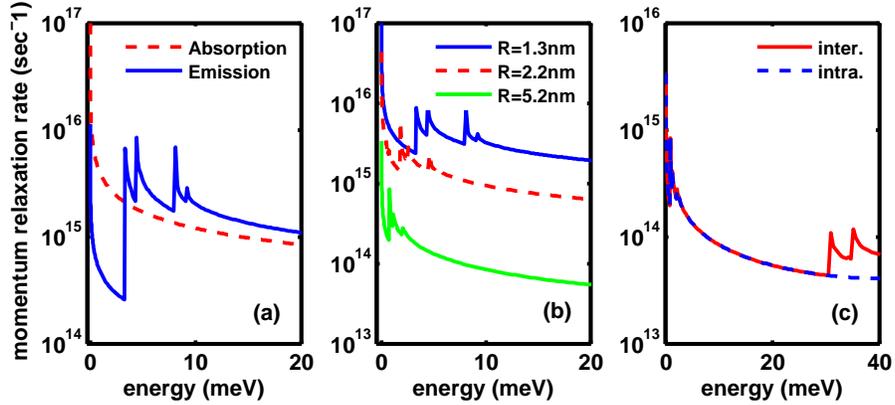}
\caption{
(a)
Momentum relaxation rate (as a function of the initial
energy of the electron) calculated for an InSb nanowire
with radius 1.3 nm.
The rates owing to phonon emission and absorption are presented
separately and only phonon emission peaks appear in the rate.
(b)
The summed total relaxation rates for
phonon emission and absorption,
for a couple of cross sectional sizes.
(c)
Effect of inclusion of the intersubband scattering
for an InSb nanowire with radius 5.2 nm (solid red curve).
Compared to the intrasubband-only-scattering (blue dashed curve),
magnitude change and additional peaks
appear in the relaxation rate of electron with
higher energies (e.g. over 30 meV in this plot).
All the results in (a), (b) and (c) are calculated
under temperature 300K.     }
\end{figure}

{\flushleft\bf Mobility.}
With the knowledge of momentum relaxation rate,
the electron mobility can be calculated within the framework
of Boltzmann equation \cite{shu12,Zim72}.
First, for the electron in the specific subband $i$,
the mobility can be calculated via
\begin{eqnarray}\label{mui}
\mu_i=\frac{2e}{k_B T m_i^*}\frac{\int_{first BZ}
[E_i(k_z)-E_0] W_i^{-1}(k_z) f_0(k_z) d k_z }
{\int_{first BZ} f_0(k_z) d k_z } \,,
\end{eqnarray}
where $m_i^*$ and $E_i$ are, respectively,
the effective mass and energy of electron of subband $i$.
$E_0$ is the lowest energy of conduction band.
The integration is over the first Brillouin zone,
and the zero-field equilibrium distribution function reads
$f_0(k_z)=e^{-E_i(k_z)/k_B T}$.
$W_i(k_z)=\sum_f W_{if}(k_z)$
is the momentum relaxation rate analyzed above.
Next, for the measured mobility in practice,
we average all the individuals simply as
\begin{eqnarray}
\mu=\frac{1}{\sum_i n_i} \sum_i{\mu_i n_i} \, ,
\end{eqnarray}
where $n_i$ is the population weight of the $i_{\rm th}$ subband.

In order to compare the results of finite size nanowire
with the bulk mobility, we quote here
the mobility formula for the three dimensional (3D) bulk systems
derived from the Boltzmann equation and under the deformation
potential approximation for the acoustic phonon scattering \cite{shu12}
\begin{equation}
\mu = \frac{2\sqrt{2\pi}\, e\hbar^4 C^{3D}}
{3 E_a^2 (m_e^*)^{5/2} (k_BT)^{3/2} }   \,,
\end{equation}
where $C^{3D}$ is the elastic constant of the 3D material
and $E_a$ is the deformation potential.
We used the InSb parameters from Ref.\ \cite{Vur01}:
$E_a=23.3$ eV, $C^{3D}=684.7$ GPa, and $m_e^*=0.0135 m_0$.

\begin{figure}[!htp]
\includegraphics[scale=0.45]{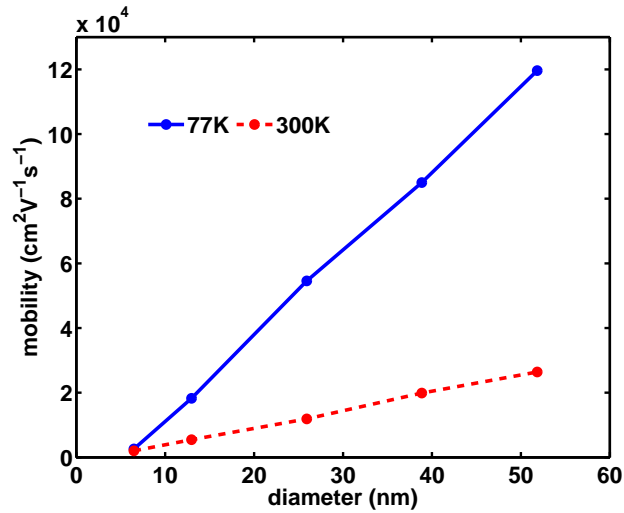}
\caption{
Size dependence of mobility of the InSb nanowires
calculated at 77K (blue dots) and 300K (red dots).   }
\label{6-mob}
\end{figure}

In our present state-of-the-art calculation, we aim at
the electron mobility of the InSb nanowires
with relatively large diameter up to 51.8 nm,
under temperatures 77K and 300K, respectively.
In this case, particular attention
should be paid to the multiple inter-subband scattering.
In our simulation, based on a self-consistent check,
we included the lowest 15 electron subbands
and 50 transverse modes of phonons.
As a comparison, we notice that in Ref.\ \cite{Ana08j,Ana08n},
for the small 3 nm diameter SiNW,
the top 3 valence subbands are taken into account
for inter-subband scattering in the hole mobility calculation,
owing to the relatively small energy spacings.

The key results are displayed in Fig.\ 3.
The easier observation is the temperature effect.
We find that only the mobility of small size nanowires
has weak dependence of temperature.
With the increase of the cross sectional size,
the temperature effect becomes more prominent,
owing to more phonons excited and involved
in the scattering which reduce then the mobility.

More complicated is the size dependence of the mobility.
First, the {\it monotonous increase} of mobility with the diameter,
which goes far beyond the size scope explored in Ref.\ \cite{Ana08j,Ana08n},
reveals different features from the small size SiNW.
In Ref.\ \cite{Ana08j,Ana08n} it was found that,
at temperature 300K, the hole mobility of the SiNW
with only up to 3.5 nm diameter is higher than the
acoustic phonon limited hole mobility of {\it bulk} Si.
This is indeed an unexpected result.
Nevertheless, at 77K,
it was found in Ref.\ \cite{Ana08j,Ana08n} that,
comparing the 3 nm diameter SiNW with bulk Si,
the hole and electron mobilities are, respectively,
8486 {\it vs} 11481 ${\rm cm}^2/{\rm V s}$ for hole,
and $\sim 2500$ {\it vs}  $23000$ ${\rm cm}^2/{\rm V s}$ for electron.

In our case, we find that up to 51.84 nm diameter
the mobility does not show a {\it saturation} behavior.
At temperature 77K, the computed value of mobility
at this size is $1.2\times 10^5\; {\rm cm}^2/{\rm V s}$,
which is lower than both
the theoretically estimated (acoustic phonon limited)
and experimentally measured values of bulk InSb mobility,
i.e., $2.8\times 10^7$ and
$1.2\times 10^6\; {\rm cm}^2/{\rm V s}$, respectively.
For temperature 300K,
the calculated mobility of the 51.84 nm diameter InSb NW
is also lower than the bulk mobility,
i.e., the theoretical (acoustic phonon limited) result
$3.6\times 10^6\; {\rm cm}^2/{\rm V s}$
and the measured value
$7.7\times 10^4\; {\rm cm}^2/{\rm V s}$.

For both temperatures, the reason that the theoretical results
are about one order of magnitude higher than the measured values
is owing to some more scattering mechanisms involved in real case.

Second, another interesting point to be noted
is that, while the bulk InSb mobility --
both the theoretical (acoustic phonon limited)
and measured values, as mentioned above --
is considerably higher than the bulk Si mobility
(e.g., at 77K, $2.3\times 10^4 {\rm cm}^2/{\rm V s}$),
but the InSb nanowire with small diameter has similar mobility
as that computed in Ref.\ \cite{Ana08j,Ana08n}
for the 3 nm SiNW, i.e., $2500 {\rm cm}^2/{\rm V s}$.
This implies that the mobility of InSb nanowire
would decrease faster than the SiNW
as we reduce the size of the nanowire.

This feature can be understood as follows.
Since the sound speed of Si material
(either the longitudinal $v_l$ or the transverse $v_t$)
is about 2.5 times faster than the one of the InSb,
we know then that the phonon frequency of large size SiNW
is roughly 2.5 times higher than the same size InSb NW
(and of course for the same wave vector).
We also know that higher frequency phonon can dissipate
carrier's energy more efficiently.
This may by one of the reasons that the bulk mobility of Si
is lower than that of the InSb.
However, with the decrease of the size of the nanowire,
the lateral confinement will generate high frequency phonons
with yet small longitudinal wave-vector.
We then expect that the mobility of the InSb NW will decease faster
than that of the SiNW with the decrease of the cross sectional size.
In particular, for such a small size of $\sim 3$ nm diameter,
both have similar magnitude.


Finally we remark that the effect of the change of the effective mass
has been taken into account in our computation through \Eq{mui}.
With the decrease of the size, the effective mass will increase gradually.
For instance, the effective masses of the lowest subbands are,
respectively, (1.025, 1.034, 1.069, 1.194, 1.5)$m_e^*$
for the diameters of (51.8, 39, 25, 12, 6)nm,
where $m_e^*$ is the bulk effective mass of InSb.
And, for a given size nanowire, the effective mass
of higher subband will increase with the subband index, c.f. Fig.\ 1(a).
From \Eq{mui} we understand that the mobility is affected
by both the scattering rate and the effective mass.
Roughly speaking, for large size nanowires,
the effective mass is not sensitive to the size,
so the mobility is largely affected by the scattering rate.
But for small size nanowires, the change of effective mass
will more severely affect the mobility.      \\
\\
\\
{\bf\large Discussion}\\
In this work we present a state-of-the-art calculation
for the phonon limited electron mobility of the InSb NWs
with diameters up to 51.8 nm.
In our calculation we carefully accounted for the
complicated intersubband electron-phonon scattering
by including the lowest 15 electron subbands
and 50 transverse modes of phonons.
These efforts extend the existing theoretical studies,
e.g., for SiNWs with small diameters about 3 nm
\cite{Ana08j,Ana08n,Fon06,Mic93,San93,San92}.
One of the reasons to make this be possible
is that, rather than the first-principle
or the tight-banding methods,
we employ the eight-band $\mathbf{k\cdot p}$
theory for the electronic structure calculation
with tolerable computational expenses.
The $\mathbf{k\cdot p}$ theory should be appropriate
for the concerned low-field mobility calculation,
for which the acoustic phonon scattering is also
the dominant intrinsic mechanism.


Applying the $\mathbf{k\cdot p}$ approach
to confined systems \cite{Pok01,Yang06,Nov05},
one should first obtain the correct $\mathbf{k\cdot p}$
Hamiltonian matrix for the bulk system, c.f. \Eq{Hnn},
which contains the {\it bulk parameters} calculated
(or extracted from measurement) at the high-symmetry point.
So the crystal variations in the structure of the InSb NWs
(e.g., the zinc-blende {\it versus} wurzite structure)
can be accounted for in this theoretical framework.
Then, the second step is to convert the wave vectors
in the cross section into momentum operators.
The new band structure is given by
the eigenvalue equation (\ref{Hnn}), 
which provides the essential information
of electronic states of the confined system.
For the calculation of phonon modes,
following the strategy in literature \cite{Ana08j,Ana08n,Yu96},
we simply adopted the DP parameter and sound speed as the bulk values.
This approximation is reasonable by noting that these parameters
rapidly tend to the bulk values with the increase of diameters,
e.g., larger than $5\sim 10$ nm \cite{Yu96,Mur10}.
Actually, this approximation has been employed also
for the very small size SiNWs \cite{Ana08j,Ana08n}
and was supported by the result of
the first-principle calculation \cite{Mur10}.

To summarize, our main result shows that,
up to relatively large diameters (e.g. 51.84 nm),
the mobility of the InSb NWs would
monotonously increases with the diameter.
This implies a remarkably smaller mobility of InSb NWs
than the bulk material.
We also find that, while the bulk InSb mobility
is considerably higher than the bulk Si,
the small size (e.g. $\sim 3$ nm diameter) InSb
and Si nanowires have similar mobilities.
This implies, importantly, that
the mobility of the InSb NWs would decrease faster than
the SiNWs as we reduce the cross sectional size of the nanowires.

The result in Fig.\ 3 is only the acoustic phonon limited mobility,
which surely overestimates the values of the mobility.
In real nanowires, beside the intrinsic phonon scattering,
there are also other scattering mechanisms to affect the mobility,
among them including such as the neutral and charged impurities
inside the nanowire and very importantly, the surface roughness
and non-idealities.
Further studies should consider these scattering mechanisms,
despite that it seems difficult to reliably model them
owing to their strong dependence on many real growth conditions.
With respect to the temperature dependence,
the thermal-excitation-related
phonon scattering should be more relevant.
Other mechanisms are insensitive to temperatures,
as long as the temperature does not affect
the impurity and roughness configurations.

Within the intrinsic mechanism of phonon scattering,
our present study provides an insight into the mobility
variation behavior of the specific InSb NWs with cross
sectional sizes, beyond most existing calculations,
despite further space remaining for future explorations.
It would be very interesting (but quite challenging)
to extend the study to larger sizes
to probe the transition to bulk behavior.
With the increase of the diameters, the huge enhancement
of both the electron and phonon subbands would make
the numerical calculations intractable.       \\
\\
\\
{\bf\large Methods}\\
\\
Under the deformation potential model
for the electron-acoustic-phonon interaction,
one can make the second quantization to the Hamiltonian
$H_{ep}=E_a\nabla \cdot \mathbf{u}$ which yields \cite{Yu96}
\begin{align}
H_{ep}
=E_a \sum_{n}\int dq_z  A_{n} (q_z^2+q_{l,n}^2)
J_0(q_{l,n} r)(\hat{a}_{n,q_z} e^{iq_zz}
+\hat{a}_{n,q_z}^{\dagger} e^{-iq_zz})  \,,
\end{align}
where $\hat{a}_{n,q_z}^\dagger$ and $\hat{a}_{n,q_z}$ are
the creation and annihilation operators of the phonon of the
$n$-th transverse mode and with longitudinal wave vector $q_z$.
The normalization factor reads
$A_{n}=\frac{R}{2}\sqrt{\frac{\hbar}{ \rho V \omega_n(q_z)S_n}}$.
$\omega_n(q_z)$ is the phonon's frequency.
$\rho$ and $V$ are, respectively, the mass density
and volume of the nanowire.
$S_n$ is given by
\begin{equation}\label{eq-Sn}
\begin{aligned}
S_n&=\frac{q_z^2R^2}{2}[J_1(q_{l,n}R)^2+J_0(q_{l,n}R)^2]
+\frac{\beta_n^2 q_{t,n}^2 R^2}{2}[J_1(q_{t,n}R)^2+J_0(q_{t,n}R)^2] \\
&+\frac{q_{l,n}^2R^2}{2}[J_1(q_{l,n}R)^2-J_0(q_{l,n}R)J_2(q_{l,n}R)]
+\frac{\beta_n^2 q_z^2 R^2}{2}[J_1(q_{t,n}R)^2-J_0(q_{t,n}R)J_2(q_{t,n}R)]  \\
&+\frac{2\beta_n q_z q_{l,n} R}{q_{l,n}^2-q_{t,n}^2}[q_{l,n}J_0(q_{l,n}R)
J_1(q_{t,n}R)-q_{t,n}J_0(q_{t,n}R)J_1(q_{l,n}R)]   \\
&-\frac{2\beta_n q_z q_{t,n} R}{q_{l,n}^2-q_{t,n}^2}[q_{t,n}J_0(q_{l,n}R)J_1(q_{t,n}R)
-q_{l,n}J_0(q_{t,n}R)J_1(q_{l,n}R)]\,.
\end{aligned}
\end{equation}
Under the free-standing boundary condition,
the coefficient $\beta_n$ in this result is given by
\begin{equation}
\beta_n=-\frac{2q_zq_{l,n}}{q_{t,n}^2-q_z^2}\frac{J_1(q_{l,n}R)}{J_1(q_{t,n}R)}.
\end{equation}

Applying the Fermi's golden rule, the rate of electron
scattered by acoustic phonons between subband states
$(i,k_z)$ and $(f,k'_z)$ reads
\begin{equation}
\begin{aligned} \label{eq-scat-1}
W_{if}(k_z,k_z')
=\frac{2\pi}{\hbar}|\langle \psi_f(k'_z)  |H_{ep}|\psi_i(k_z)\rangle|^2
\delta[E_f(k_z')-E_i(k_z)+\epsilon \hbar \omega_n(q_z)] \,,
\end{aligned}
\end{equation}
where $k_z$ and $k'_z$ are, respectively, the initial
and finial momentum of the electron.
$\epsilon=\pm 1$ correspond to emission and absorption
of a phonon (with momentum $q_z$),
while the condition of energy conservation is implied
by the $\delta$ function.

In \Eq{eq-scat-1}, the scattering matrix element is detailed as
\begin{equation} \label{H-ep}
\begin{aligned}
\langle \psi_f(k'_z)  |H_{ep}|\psi_i(k_z)  \rangle
=E_a \sum_n \sum_{\epsilon=\pm 1}
\int d q_z  A_{n} \frac{[\omega_n(q_z)]^2}{v_l^2} \frac{2\pi}{l} \delta(k_z-k_z'-\epsilon q_z)
\, F_{if}(q_{l,n})\hat{a}_{n,q_z}^\epsilon \,.
\end{aligned}
\end{equation}
Here we have introduced $\hat{a}_{n,q_z}^{\pm 1}$ to denote
$\hat{a}^{\dagger}$ and $\hat{a}$, respectively.
We also introduced the reduced overlap integral as
\begin{equation}
\begin{aligned}
&F_{if}(q_{l,n})=
\frac{2}{J_{L+1}(\alpha^L_m) J_{L'+1}(\alpha^{L'}_{m'})}
\int^1_0 d (\frac{r}{R})(\frac{r}{R}) J_{L}
(\frac{\alpha^{L}_{m}r}{R})J_{L'}
(\frac{\alpha^{L'}_{m'}r}{R})J_0(q_{l,n} r)\,.
\end{aligned}
\end{equation}
The $\delta$-function in \Eq{H-ep}
reflects the momentum conservation
in the axial direction of the nanowire,
while in deriving this result we used
the method of {\it box normalization}, i.e.,
$\delta(k'-k)=\delta(0)=l/(2\pi)$ when $k'=k$.

Now, applying the formula
$\delta[\phi(x)]=\sum_k \frac{\delta(x-x_k)}{\phi'(x_k)}$,
we transform the $\delta$-function of energy conservation
in \Eq{eq-scat-1} as follows:
\begin{equation}
\begin{aligned}
&\delta[E_f(k_z')-E_{i}(k_z)+\epsilon \hbar \omega_n(q_z)]  \\
&=\sum_p \delta(q_z-\tilde{q}_{z,p})/|\partial [E_f(k_z+\epsilon \tilde{q}_{z,p})+\epsilon \hbar \omega_n(\tilde{q}_{z,p})]/\partial q_z  | \,. \\
&=\sum_p \delta(q_z-\tilde{q}_{z,p})JDOS(k_z,\tilde{q}_{z,p}) \,.
\end{aligned}
\end{equation}
Here we have introduced the {\it joint density of states}
of electron and phonon as
\bea
JDOS(k_z,\tilde{q}_{z,p})
=|\partial [E_f(k_z+\epsilon \tilde{q}_{z,p})
+\epsilon \hbar \omega_n(\tilde{q}_{z,p})]/\partial q_z|^{-1}  \,.
\eea
In the above two equations,
$\tilde{q}_{z,p}$ is the root of the algebraic equation
$E_f(k_z')-E_{i}(k_z)\pm \hbar \omega_n(q_z)=0$.

After these preparations, we obtain the final expression
we used in this work for the momentum relaxation rate as
\begin{equation}\label{eq-mom}
\begin{aligned}
&W_{if}(k_z)=\sum_{n,p,\epsilon=\pm 1}
\int dk_z' W_{if}(k_z,k_z',q_z,n) (1-\frac{k_z'}{k_z})   \\
&=\frac{E_a^2}{4\pi  \rho v_l^4}
\sum_{n,p,\epsilon} S_n^{-1}\frac{\epsilon \tilde{q}_{z,p}}{k_z}
|F_{if}(q_{l,n})|^2 [\omega_n(\tilde{q}_{z,p})]^3
\, JDOS(k_z,\tilde{q}_{z,p})
\, [N_{\rm ph}(\omega_n(\tilde{q}_{z,p}))+(1+\epsilon)/2] \,,
\end{aligned}
\end{equation}
$N_{\rm ph}$ is the average number of the thermal phonons,
with frequency $\omega_n(\tilde{q}_{z,p})$.
The weighting factor $(1-k'_z/k_z)$
is from a consideration
of angle dependence of the scattering.
That is, in the case of $k'_z<k_z$,
the electron either emits a phonon with $q_z>0$
or absorbs a phonon with $q_z<0$, resulting thus
in an increase of the momentum relaxation rate
and decrease of the electron mobility.
On the contrary, if $k'_z>k_z$,
the momentum relaxation rate is reduced.
Finally, we mention that
in present study we omitted the Umklapp process.


\vspace{0.5cm}
{\flushleft\bf\large Acknowledgements}\\
This work was supported by the Major State `973'
Project of China (No.\ 2012CB932704)
and the NNSF of China (Nos. 11675016 \& 21421003).

\vspace{0.2cm}
{\flushleft\bf\large Author contributions} \\
X.Q.L. and W.F. designed the work;
W.F., C.P. and S.L. carried out the calculations;
X.Q.L. participated in all the details and
wrote the paper and all authors reviewed it.

\vspace{0.2cm}
{\flushleft\bf\large Additional information}  \\
Competing financial interests:
The authors declare no competing financial interests.

\end{document}